# The Progress of Astronomy Program in the Rizal Technological University Philippines

**Ryan Manuel D. Guido**
*Program Officer, Center for Astronomy Research and Development, Rizal Technological University, Mandaluyong City, Philippines*

**Abstract.** The Rizal Technological University offers a formal degree program in Astronomy for both undergraduate and graduate level. Also, the university hosts the most sophisticated optical telescopes in the country, and the astronomical research hub Center for Astronomy Research and Development. Through the efforts of the Department of Science and Technology, it offers support for scholarships and research funding. The astronomy curriculum of the university aimed to craft students to become researchers and scientists in the future, and notably associates the tracer study that the graduates are equipped as researchers and possesses the computational and mathematical ability for the job. The increased in enrollees shows the cogency of the aforecited statements as it enticed students to study in the field of astronomy. Other efforts are discussed in detail in this paper.

## *1. Introduction*

Several years have passed, the Philippines have established astronomy education and research. It may still be behind with neighboring Southeast Asian countries, especially those who have had established far way back and has been established advanced space technologies. At that time, it was very evident that there was the scarcity of trained professional Filipino astronomers (Kouwenhoven & Sese, 2012). Astronomy in the Philippines started with the efforts of amateur astronomers wherein they were able to showcase to Filipino community observational astronomy and let pedestrians experience the admiration of the celestial beauty.

With the desire of the Filipino people to explore the cosmos, they have initiated several outreach and stargazing activities. Through the help of the Department of Science and Technology (DOST) Philippine Atmospheric Geophysical, Astronomical Services Administration (PAGASA) led the frontier endeavors of astronomy in the Philippines. Elevating astronomy and space science teaching preparation has been the main focus of educational institutions for the past decade.

## *2. History of Astronomy in the Philippines*

The Manila Observatory was established in 1865 (formerly known as Observatorio Meteorologica De Manila) by Jesuits Scientist. It was originally built in the suburbs of Intramuros Manila. It is now named after Padre Faura – the founder of the observatory. It was engaged in weather observations and forecasting but later expanded into the study of meteorology, astronomy, and geomagnetism. It was then established by the American colonial government as the Philippine Weather Bureau (Torres, 2007). The astronomical practices are mainly restricted towards delivering timekeeping, solar and stellar observation (Celebre, 2008). Initially operated as a private institution in 1865, it has become a government agency in 1901 as the Weather Bureau, and at present time is the well-known PAGASA (Soriano, 1996). In 1963, the Manila Observatory relocated to Loyola Heights campus of the Ateneo de Manila University, where it continued it's seismic, geomagnetic, and radiophysics research (observatory.ph).

At present, the main institutions in the Philippines that perform astronomical functions are the DOST – PAGASA, the National Planetarium (NP), the Manila Observatory (MO), the Rizal Technological University (RTU), New Era University (NEU).

The endeavors of PAGASA has brought the Philippines to higher grounds of international recognition with the inclusion of the Philippines as Associate Member of the International Astronomical Union (IAU) in 2000. It also led the promotion of astronomy as well as space science by the virtue of Presidential Proclamation 130, otherwise known as the celebration of the National Astronomy Week (NAW) every third week of February. It aims to carry out activities such as; astronomy lectures, stargazing and telescope sessions and mobile planetarium shows (Celebre, 2008).

*3. Astronomical Resources*

In recent years there are just two institutions with telescopes: The University of the Philippines (UP) and the Ateneo de Manila University, which has the Manila Observatory in its compound. The Japanese government donated a 400-mm observatory-type telescope of the Cassegrain to UP. It is the second largest telescope in the country in terms of aperture. The biggest telescope in the Philippines is the 450-mm Cassegrain, which is hosted in the Astronomical Observatory of the PAGASA. This telescope was likewise given by the Japanese government. The 400-mm at the UP is regularly utilized by students during astronomy-related activities, and the 450-mm at the PAGASA is utilized for the most part for CCD imaging of variable stars. The Manila Observatory has a variety of equipment for solar and atmospheric observation (Torres, 2004).

Today, RTU hosts the most sophisticated optical astronomy equipment in the country. It has the most advanced observational astronomy equipment which can deliver highly competitive astronomy education and research. Despite the university is located in the urban area, it has embraced the challenges of light pollution in the city and has brought up the first research study of light pollution in the country.

The RTU now houses the Center for Astronomy Research and Development which is the hub for astronomy and space science technology research and development in the country. Through this research and development effort, it will cater to the enticement of young scientists in the field, by conducting capacity-building activities for astronomy researchers.

The Center has several research laboratories; Light Pollution Research Laboratory (LiPReL); Astronomical Instrumentation and Manufacturing Research Laboratory (AIMReL); Atmospheric & Space Weather Research Laboratory (AtReL); Astronomy and Astrophysics Symbolic Computational Research Laboratory (ASCReL); Astronomy and Space Technology Research Laboratory (ASTReL); and Astronomy for the Public Research Studies (APReS).

*4. Astronomy Education*

In the past, astronomy is taught as a fragment of the general science subject in basic education. To elevate the context of astronomy education, the National Institute of Physics (NIP) of the University of the Philippines (UP) offered the subject "Physics and Astronomy for Pedestrians" in 2002 (Celebre, 2008). It deals with the introduction to the different aspects of physics and astronomy, from its emergence up to its current developments (Celebre, 2003).

Lecturers took the course on Astronomy and Astronomical Observation at GAO in 2001, and batches of science personnel participated in the International School for Young Astronomers (ISYA), the signing of Memorandum of Agreement between IAU/TAD and PAGASA in 2002 for the conduct of Astronomers Training Course (Celebre, 2008).

In 2005, the RTU marked a great revolution in the history of Philippine education in the field of astronomy, the university offered the very first formal degree program of Master of Science in Astronomy (Celebre, 2008; Torres, 2007). Its curriculum is designed to students with any bachelors' degree. Furthermore, RTU took another leap by its undisputed leadership in astronomy education as it offered the Bachelor of Science in Astronomy Technology (Celebre, 2007). It is a collaborative foundation of RTU and PAGASA (Torres, 2007). It is customized with the combination of pure astronomical science and engineering technology.

*4.1 The Astronomy Curriculum*

The BS Astronomy Technology curriculum is a program designed to study astronomy in various configurations with the inclusion of technological components to ensure the ability of the graduates to contribute to the growth of science in the country. The technology was integrated into the curriculum as RTU as one of the leading universities in the Philippines int eh field of engineering and technology. With the aid of the expertise of entrepreneurs, scientists, engineers, and the academe, they supported that the program will lead the graduates in scientific careers in government or research agencies (Torres, 2007). After eight (8) years in 2015, New Era University (NEU) followed the step of RTU and has become the first private higher education institution in the Philippines to offer BS Astronomy program.

In 2018, RTU has undergone revision of the curriculum with the support from its Astronomy Advisory Council, which aims to elevate the position of astronomy education in the country. To cater the needs of the scientific community in the Philippines, the RTU develops astronomy curriculum which offers Astrophysics, Meteorological Sciences, and Space Science Technology as part of their academic tracks. It will lead to producing significant human resources suitable to the needs of the fourth industrial revolution specifically in the field of space science.

Students pursuing an astronomy degree should have a sufficient background in science and mathematics. It is very evident that the foundations of studying astronomy is dealing with physics. Physics is pondered as one of the most prevalent and challenging subjects by the students in the realm of science (Guido, 2013). Thus, it is suggested that students to succeed in classes is to impart invaluable educational teaching techniques that build a framework in the evaluation of the students' cognitive ability, behavior, and attitudes of learners (Guido, 2014).

As physics conception is being one of the criteria to accept students, a physics diagnostic test is being administered to accept students in the department besides the usual college entrance exam. It has been found out that it is a relevant approach to qualify astronomy students (Guido & Dela Cruz, 2015). With this process, students are then prepared in mind the challenges they have to face in studying astronomy. It showed that students are optimistic in their academic performance, self-reliant with their study habits, confident with their learning styles, and ardent in their drive to learn (Guido & Dela Cruz, 2013). This also reveals that professors should also provide more vivid teaching strategies to the students that can cope up with their studies (Guido, 2015).

With the passing rate of about 65%, the RTU entrance examination is one of the toughest entrance exams in the region. With five hundred twenty (520) student applicants for BS Astronomy, only one hundred eighty (180) successfully passed the exam. This uncovers that students who passed the exam are genuinely qualified and have sufficient knowledge to undertake the program.

**Table 1. Credit units of curricular domains of the BS Astronomy Program**

| Curricular domains | Units | Fields or Subjects |
|---|---|---|
| General Education | 30 | Understanding Self, Readings in Philippine History, The Contemporary World, Mathematics in the Modern World, Purposive Communication, Art Appreciation, Science, Technology and Society, Ethics, Life and Works of Rizal, Environmental Science |
| Languages | 12 | Fil01, Fil02, Fil03, Eng01 |
| Physical Education | 8 | PE01, PE02, PE03, PE04 |
| National Service Training Program | 6 | NSTP1, NSTP2 |
| Chemistry | 4 | Applied Inorganic and Organic Chemistry Lecture and Laboratory |
| Physics | 17 | Mechanics, Mechanics Laboratory, Electromagnetics and Waves, Optics and Designs with Instrumentation, Optics and Designs with Instrumentation Laboratory, Modern Physics, Quantum Particle Physics |
| Technology | 1 | Computer Aided Drafting (Aircraft and Satellite Designs) |
| Mathematics | 9 | Pre-Calculus, Differential Calculus, Integral Calculus, Differential Equations |
| Cognates | 6 | Industrial Organizational Management, Entrepreneurship, Cooperative Education (Practicum) |
| Astronomy | 75 | Introduction to Astronomy, Planetary Science, Astrochemistry, Observational Astronomy I & II Lec and Lab, Astronomy Instrumentation, Astronomical Programming I & II Lec and Lab, Astronomical Research Laboratory Immersion I & II, Data Science and Analysis, Astrobiology and Life in the Universe, Celestial Mechanics, Astrophysics, Astronomy Research I & II Lec and Lab, Meteorology, Scientific Paper Writing I & II, Material Science, Archeaoastronomy and Ethnoastronomy, Seminar in Astronomy, Astronomy Education, Planetarium & Observatory Design Lec and Lab, Community Based Astronomy |
| Major | 12 | Remote Sensing, Space Physics, Stellar Astrophysics, Galactic and Extragalactic Astrophysics |

*Passing the comprehensive examination is required to proceed to third year**

With about one hundred eighty (180) units to complete the program in four (4) years, about fifty (50%) percent is pure astronomy while the remaining half are subjects being mandated by the Philippine government to be taken by the students.

### 4.2 Practicum

The Practicum of the students will be held at the National Astronomical Research Institute of Thailand and the Chiang Mai University wherein RTU has international linkages and inks collaborative efforts to train faculty and students in the field of astronomy and space science. The International practicum will instill significant and relevant researches wherein the student will experience doing researches at laboratories abroad. In the succeeding years, the university will benchmark and link ties to other international astronomical research institutions or space agencies to broaden the relations and skills development of the students.

### 4.3 Scholarship Program

Since NEU is a private institution, they charged students with a tuition fee of about twenty thousand pesos (P 20,000) per semester. While RTU is a state university, it is covered with the Republic Act 10931 or the University Access to Quality Tertiary Education Act (UAQTEA), which mandates the free tuition and miscellaneous fees in state universities and colleges (SUCs). This implies that enrolled college students need not to pay for their tuition, miscellaneous, and other similar or related fees. An initial budget of forty billion (P 40,000,000) pesos was allocated by the government for this purpose.

The BS Astronomy program is also recognized by the Department of Science and Technology Science Education Institute (DOST – SEI) under the Science and Technology Undergraduate Scholarships and Junior Level Science Scholarships as one of their priority courses. The Science and Technology Undergraduate Scholarship aims to entice and encourage brilliant Filipino youths to pursue lifetime dynamic careers in science and technology and ensure a sound, sufficient resource of qualified S&T human capital that can lead towards the progress of the country.

Student scholarships can be classified under Republic Act 7687 (RA 7687) otherwise known as the "Science and Technology Scholarship Act of 1994" this provides students a scholarship to talented and eligible students whose families' socio-economic status does not exceed the set cut-off values of certain indicators. While the DOST-SEI Merit Scholarship Program, formerly known as the NSDB or NSTA Scholarship under RA No 2067 is awarded to students with high aptitude in science and mathematics and is willing to pursue careers in the fields of science and technology. And Republic Act 10612 (RA 10612) aims to strengthen the country's science and technology education by fast-tracking graduates in the sciences, mathematics, and engineering who shall teach science and mathematics in the secondary schools throughout the country.

The student who successfully availed of the scholarship has the following privileges:

Table 2. DOST Scholarship privileges

| Privileges | Amount |
| --- | --- |
| Tuition and other school fees | P 40,000/ yr |
| Book Allowance | P 10,000/ yr |
| Uniform | P  1,000/ yr |
| Group Health and Accident Insurance | Premium |
| Transportation allowance (for those studying outside of home province) | 1 economy-class roundtrip fare |
| Monthly living allowance | P  7,000/ month |
| Graduation Clothing Allowance | P  1,000 |

### 4.4 Astronomy Enrollees

Being the hub for astronomy education in the Philippines, and with an increasing number of students being motivated to pursue degree programs in astronomy. The table and figure below show it has earned a thirty-five (35%) percent upsurge in enrollees per year.

Table 3. Enrollment Summary of the BS Astronomy Program

| School Year Enrollment | Freshmen | Higher years | Total |
| --- | --- | --- | --- |
| 2011 - 2012 | 41 | 36 | 77 |
| 2012 - 2013 | 47 | 61 | 108 |
| 2013 - 2014 | 45 | 80 | 125 |
| 2014 – 2015 | 50 | 104 | 154 |
| **2015 - 2016** | 44 | 119 | 163 |
| **2016 – 2017** | *No new enrollees due to implementation of the* | | |
| **2017 – 2018** | *Senior High School Grade 11 and Grade 12* | | |
| **2018 – 2019** | 166 | 68 | 234 |
| **2019 - 2020** | 191 | 133 | 324 |

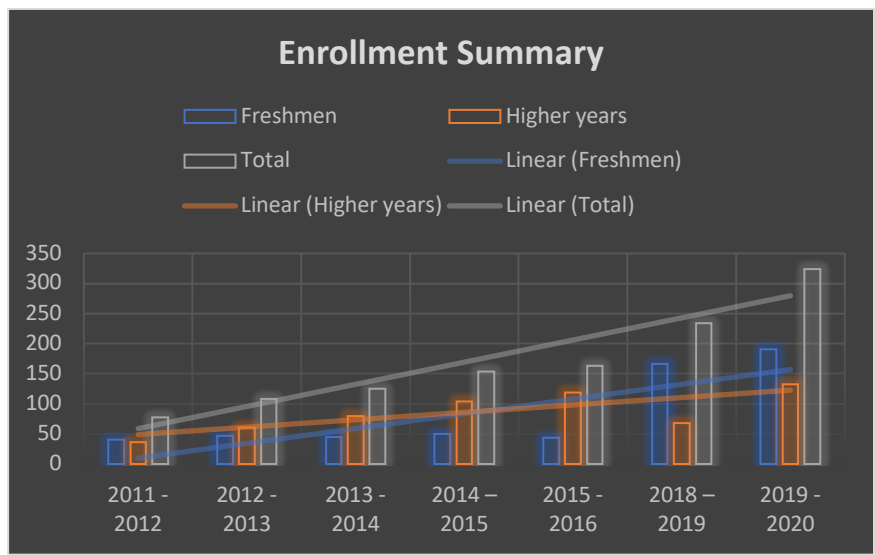

Figure 1. Enrollment Summary

With this growth, and with a limited number of resources such as classrooms, research laboratories, and faculty members; the expected freshmen enrollees will be limited to two hundred (200), and higher years will be two hundred ninety-four (294), totaling to 494 student pollution by the school year 2020 -2021.

*4.5 Tracer Study*

The BS Astronomy graduates have a strong passion for the profession, and the graduate's experiences in the course study is the development of astronomy in the Philippines. Most of the graduates are employed and currently working in private and public national government. It only took them at most of six (6) months to get a job. About half of the graduates pursue graduate studies, some of whom are full-time and most are part-time students. Their mean salary ranges from Php 10,000 – 20,000. The most dominant skills their employers admired is their investigative research skills, followed by computational and mathematical skills, while oral communication skills are identified to be of improvement (Miranda, 2017).

*5. Space Programs in the Philippines*

The PHL Microsat Program was established to build, launch and effectively utilized micro-satellite technology for multispectral, high precision earth observation. The Diwata satellites aim to render scientific earth observations capabilities for resource and disaster risk reduction management, resource assessment, environmental monitoring and weather observations in the Philippines through the efforts of Filipino and Japanese Scientists.

The PHL Microsat Program is now succeeded into Space Technology and Applications Mastery, Innovation and Advancement (STAMINA4Space) program. It is funded by the Department of Science and Technology, monitored by the DOST Philippine Council for Industry, Energy and Emerging Technology Research and Development (DOST-PCIEERD) and jointly implemented by the University of the Philippines – Diliman and DOST – Advanced Science and Technology Institute and Hokkaido University and Tohoku University with related academe-industry-government partnerships.

The DOST has formulated the National Space Development and Utilization Policy (NSDUP). It focuses on six (6) key development areas (KDA) to develop space science and technology in the country (i) National Security and Development, (ii) Hazard Management and Climate Studies, (iii) Space Research and Development (iv) Space Industry Capacity Building, (v) Space Education and Awareness, and (vi) International Cooperation. The KDAs aims to preserve and enhance the country's national security and promote development beneficial to all Filipinos, to develop and utilize space technology application (STA) to enhance its hazard management and disaster mitigation strategy and ensure the nation's resilience to climate change. It also aims to spur rapid scientific growth, the Philippines focus on conducting R&D endeavors in vital areas of space science, technology, and allied fields. Likewise, it intends to create a robust and thriving space industry to support the country's space program through private sector involvement and cooperation. Moreover, the development of expert manpower, curriculum and education materials at all levels, as well as to become a key player in the ASEAN and global space community as a service and manpower-provider are the goals of the DOST.

The Philippine Space Agency (PhilSA) was put into law through Republic Act 11363 known as establishing the Philippine Space Development and Utilization Policy and creating the Philippine Space Agency and for other purposes signed last 08 August 2019. It is the country's strategic roadmap for space development to become a space-capable and space-faring nation. It focuses on space science technology application to address national issues on defense, assets and resources, disaster risks, and other commitments. It will exercise its functions on (i) policy, planning and coordination, (ii) improved public access and resource sharing, (iii) research and development (iv) education and capacity building, (v) industry development, and (vi) international cooperation.

## 6. Conclusion

Astronomy in the Philippines now has direction. Soon, with the help of the government and Filipino scientists as human resources, the country will eventually mark its pathway in traversing noteworthy astronomy and space technology endeavors for the Filipino people that is competitively prepared for the fourth industrial revolution. Recruiting competitive faculty members is one of the critical factors in improving and sustaining the quality of astronomy education in the country. With the emergent of science in this era, continuous retraining should be undertaken by the faculty members to be aware of urgent issues, technological advancement, and the latest astronomical discoveries.

With the continuous support of the government from its free tuition, to expanded scholarships, and its space programs, the students are getting more motivated to study by learning that space programs in the Philippines are sustainable and there is a career that waits for them in the future.

The establishment of the formal degree offerings in astronomy at Rizal Technological University emphasizes on the capacity of the country to produce scientific career opportunities and delve into a scientific field where a developing country can be competitive.

**References**

Celebre, C.P. 2007. Astronomy in the Philippines, A Hundred Years After. Thailand National Astronomy Meeting. Bangkok, Thailand.

Celebre, C.P. 2003. The Establishment of an Astrophysics Course in the Philippines. International Astronomical Union Asia Pacific Meeting.

Celebre, C.P. and B.M. Soriano, Jr. 2000. Revitalizing Astronomy in the Philippines. Published in the Astronomy for Developing Countries, proceedings of a Special Session of the International Astronomical Union, Manchester, United Kingdom, August 2000.

Guido, R.M., & Dela Cruz, R. 2015. Physics Diagnostic Test: An Approach to Qualify Astronomy Students. International Letters of Social and Humanistic Sciences. (63), 129-135.

Guido, R.M. 2015. Performance of Students in the Departmental Examination in Chemistry. International Letters of Social and Humanistic Sciences. (55), 121-125.

Guido, R.M. 2014. Development and Evaluation of Instructional Materials in Material Science and Engineering. American Journal of Educational Research, 2 (11), 1126- 1130.

Guido, R.M., & Dela Cruz, R. 2013. Factors Affecting Academic Performance of BS Astronomy Technology Students. International Journal of Engineering Research and Technology (IJERT). 2 (12), 84-94.

Guido, R.M. 2013. Attitudes and Motivation towards Learning Physics. International Journal of Engineering Research and Technology (IJERT). 2 (11), 2087-2093.

Kouwenhoven, M.B.N, and Sese, R.M.D. 2012. Developing Astronomy Research and Education in the Philippines. International Astronomical Union. Doi:10.1017/S1743921314012198

Soriano, Jr., B.M. and Celebre, C.P. 1996. Astronomy in the Philippines, UN guidebook on Developing Astronomy and Space Science Worldwide. Germany, 1996.

Torres, J. 2004. Urban Astronomy in the Philippines. Astronomy Education Review. 3 (1) 115 – 124.

Torres, J. 2007. Bachelor of Science in Astronomy Technology: A Model. Astronomy Education Review. 6 (1) 163 – 170.